\documentclass{article}
\usepackage{amssymb}
\usepackage{amsfonts}
\usepackage{amsmath}
\usepackage{mathrsfs}
\usepackage{graphicx}%
\usepackage{subfigure}
\usepackage{xcolor}
\usepackage{multirow}
\usepackage{caption} 
\definecolor{IgorGreen}{RGB}{76,153,0}
\definecolor{IgorBlue}{RGB}{0,0,204}
\definecolor{IgorGreen1}{RGB}{0,153,77}
\definecolor{IgorPurple}{RGB}{103,0,204}
\usepackage{flushend,cuted}
\usepackage[noblocks]{authblk}

\providecommand{\U}[1]{\protect\rule{.1in}{.1in}}
\begin{document}
%
\title{The Accuracy of Mean-Field Approximation for Susceptible-Infected-Susceptible Epidemic Spreading}
\author{B. Qu\thanks{ Faculty of Electrical Engineering, Mathematics and Computer
Science, Delft University of Technology, The Netherlands; \emph{email}:
\{B.Qu,H.Wang\}@tudelft.nl } \ and H. Wang}

\maketitle

\begin{abstract}
The epidemic spreading over a network has been studied for years by applying the mean-field approach in both homogeneous case, where each node may get infected by an infected neighbor with the same rate, and heterogeneous case, where the infection rates between different pairs of nodes are also different.
Researchers have discussed whether the mean-field approaches could accurately describe the epidemic spreading for the homogeneous cases but not for the heterogeneous cases.
In this paper, we explore if and under what conditions the mean-field approach could perform well when the infection rates are heterogeneous.
In particular, we employ the Susceptible-Infected-Susceptible (SIS) model and compare the average fraction of infected nodes in the metastable state, where the fraction of infected nodes remains stable for a long time, obtained by the continuous-time simulation and the mean-field approximation. We concentrate on an individual-based mean-field approximation called the N-intertwined Mean Field Approximation (NIMFA), which is an advanced approach considered the underlying network topology.
Moreover, for the heterogeneity of the infection rates, we consider not only the independent and identically distributed (i.i.d.) infection rate but also the infection rate correlated with the degree of the two end nodes.
We conclude that NIMFA is generally more accurate when the prevalence of the epidemic is higher. Given the same effective infection rate, NIMFA is less accurate when the variance of the i.i.d.\ infection rate or the correlation between the infection rate and the nodal degree leads to a lower prevalence.
Moreover, given the same actual prevalence, NIMFA performs better in the cases: 1) when the variance of the i.i.d.\ infection rates is smaller (while the average is unchanged); 2) when the correlation between the infection rate and the nodal degree is positive.
Our work suggests the conditions when the mean-field approach, in particular NIMFA, is more accurate in the approximation of the SIS epidemic with heterogeneous infection rates.   

\end{abstract}

\section{Introduction}
\label{Intro}
By considering the system components or individuals as nodes and the interactions or relations in between nodes as links, networks have been used to describe the biological, social and communication systems. On such networks or complex systems, viral spreading models have been used to describe processes e.g.\ epidemic spreading and information propagation \cite{daley2001epidemic,pastor2001epidemic,wang2013effect,0295-5075-105-6-68004,liu2015epidemics}. The Susceptible-Infected-Susceptible (SIS) model is one of the most studied models. In the SIS model, at any time $t$, the state of a node is a Bernoulli random variable, where $X_i(t)=0$ represents that node $i$ is susceptible and $X_i(t)=1$ if it is infected. Each infected node infects each of its susceptible neighbors with an infection rate $\beta$. The infected node can be recovered with a recovery rate $\delta$. Both infection and recovery processes are independent Poisson processes. The ratio $\tau\triangleq\beta/\delta$ is called effective infection rate, and when $\tau$ is larger than the epidemic threshold $\tau_c$, the epidemic spreads out with a nonzero fraction of infected nodes in the metastable state. The average fraction of infected nodes $y_\infty$ in the metastable state, ranging in $[0,1]$, indicates how severe the influence of the virus is: the larger the fraction $y_\infty$ is, the more severely the network is infected.

In this paper, we concentrate on deriving the average fraction $y_\infty$ of infected nodes in the metastable state. Although the continuous-time Markov theory can be used to obtain the exact value of $y_\infty$, the number of states is too large to be solved in a large network \cite{pastor2014epidemic}. Hence, the derivation of the average fraction $y_\infty$ of infected nodes in the metastable state mostly relies on different kinds of mean-field theoretical approaches. The first approach to study the SIS model in complex networks is a degree-based mean-field (DBMF) theory, also called heterogeneous mean-field (HMF) approximation, proposed by Pastor-Satorras et al.~\cite{Pastor-Satorras2001}, which assumes that all nodes with the same degree are statistically equivalent, i.e.\ the infection probabilities of those nodes are the same. An individual-based mean-field (IBMF) approximations, called the N-Intertwined Mean-Field Approximation (NIMFA), of the SIS model is then introduced \cite{van2009virus} with the only assumption that the state of neighboring nodes is statistically independent.
A few extensions of the above DBMF and IBMF theories are also developed \cite{eames2002modeling,gleeson2011high,boguna2013nature,mata2013pair}.
NIMFA, taking the network topology into account, turns out to be more precise on different types of networks for the classic SIS model with the homogeneous infection rates\cite{li2012susceptible} while comparing to the DBMF approximation. 
However, as discussed in \cite{qu2015sis,buono2013slow,fu2008epidemic,yang2012epidemic}, the infection rates could be heterogeneous, i.e.\ the infection rates between different pairs of nodes could also be different.
The accuracy of NIMFA, when the infection rates are heterogeneous, has not yet been discussed.

In this paper, we explore the influence of the heterogeneous infection rates on the precision of NIMFA.
In particular, we compare the average fraction $y_\infty$ of infected nodes as a function of the effective infection rate $\tau$ computed by NIMFA to that obtained by the continuous-time simulations of the exact SIS model when the infection rates are heterogeneous but the recovery rate is the same for all nodes.
In fact, the effective infection rate $\tau$ refers to the average infection rate divided by the recovery rate in the SIS model with heterogeneous infection rates.
We set the average infection rate to $1$ and tune the recovery rate $\delta$ to control the effective infection rate $\tau$.
We consider both the independent and identically distributed (i.i.d.) and the correlated heterogeneous infection rates in different network topologies.
For the case of i.i.d.\ infection rates, we employ the log-normal distribution to generate the infection rates as in \cite{qu2015sis}.
In this case, we tune the variance of the infection rates and explore when NIMFA performs better, i.e.\ the average fraction of infected nodes obtained by NIMFA is closer to that by the continuous-time simulations. 
For the case of correlated infection rates, we assume that the infection rate $\beta_{ij} (=\beta_{ji})$ between node $i$ and $j$ is correlated with their degrees $d_i$ and $d_j$ in the way: 
\begin{equation}
\label{eq1}
\beta_{ij}\sim(d_id_j)^{\alpha}
\end{equation} 
and $\alpha$ indicates the strength of the correlation.
As discussed in \cite{qu2016heterogeneous}, such a correlation between the infection rate and the nodal degree is motivated by the real-world datasets.
Moreover, the correlation strength $\alpha\approx 0.5$ in the network of airports (both in US \cite{barrat2004architecture,macdonald2005minimum} and China \cite{li2004statistical}) and $\alpha\approx 0.8$ in the metabolic network \cite{macdonald2005minimum}. 
Given a network, when we generate the heterogeneous infection rates as (\ref{eq1}), the distribution of infection rates actually changes with the parameter $\alpha$, although the average infection rate is kept to be the constant $1$.
In the case of correlated infection rates, we consider as well the corresponding uncorrelated heterogeneous infection rates scenario, where the correlated infection rates are shuffled and randomly assigned to all the links as a reference scenario, so that we can explore how the correlation between the infection rate and the nodal degree influence the accuracy of NIMFA.    

\section{Preliminary}
In this section, we introduce the foundation of this paper, including the construction of network models, the i.i.d.\ heterogeneous infection rates, the correlated heterogeneous infection rates, the mean-field approximation of the SIS model (NIMFA) and the continuous-time simulation set up.
\subsection{Network models}
\label{sec_top}
The scale-free (SF) model has been used to capture the scale-free nature of degree distribution in real-world networks such as the Internet \cite{caldarelli2000fractal} and World Wide Web \cite{albert1999internet}: $\text{Pr}[D=d]\sim d^{-\lambda}, d\in[d_{\rm min},d_{\rm max}]$, where $d_{\rm min}$ is the smallest degree,
$d_{\rm max}$ is the degree cutoff, and $\lambda>0$ is the exponent
characterizing the broadness of the distribution
\cite{barabasi1999emergence}. In real-word networks, the exponent $\lambda$ is usually in the range $[2,3]$, thus we confine the exponent $\lambda=2.5$ in this paper. We further employ the smallest degree $d_{\rm min}=2$, the natural degree cutoff $d_{\rm max}=\lfloor N^{1/(\lambda-1)}\rfloor$ as in \cite{PhysRevLett.85.4626}, and the size $N=1000$. Hence, the average degree is approximately $4$. 

The Erd\"os-R\'enyi (ER) model \cite{erdds1959random} has also been taken into account. In an ER random network with $N$ nodes, each pair of nodes is connected with a probability $p$ independent of the connection of any other pair. The distribution of the degree of a random node is binomial: $\text{Pr}[D=d]=\binom{N-1}{d}p^d(1-p)^{N-1-d}$ and the average degree is $E[D]=(N-1)p$. For a large $N$ and a constant $E[D]$, the degree distribution is Poissonian: $\text{Pr}[D=d]=e^{-E[D]}E[D] ^{d}/d!$. We consider the ER networks with the size $N=1000$ and the average degree $E[D]=4$.

\subsection{The N-Intertwined Mean-Field Approximation of the SIS model}
The N-Intertwined Mean-Field Approximation (NIMFA) is so far one of the most accurate approximations of the SIS model that takes into account the influence of the network topology. For the classic SIS model with the homogeneous infection rate $\beta$ and recovery rate $\delta$. The single governing equation for a node $i$ in NIMFA is 
\begin{equation}
\label{equ_single_NIMFA}
\frac{\mathrm{d} v_i(t)}{\mathrm{d} t}=-\delta v_i(t)+\beta(1-v_i(t))\sum_{j=1}^{N}a_{ij}v_{j}(t)
\end{equation}
where $v_i(t)$ is the infection probability of node $i$ at time $t$, and $a_{ij}=1$ or $0$ denotes if there is a link or not between node $i$ and node $j$. With $V(t)=[v_1(t)~v_2(t)~\cdot\cdot\cdot~v_N(t)]^T$, the matrix evolution equation of NIFMA is 
\begin{equation}
\frac{\mathrm{d} V(t)}{\mathrm{d} t}=(\beta\text{diag}(1-v_i(t))A-\delta I)V(t)
\end{equation}
where $A$ is the $N\times N$ adjacency matrix of the network with elements $\alpha_{ij}$, $I$ is the $N\times N$ identity matrix and diag$(v_i(t))$ is the diagonal matrix with elements $v_1(t),v_2(t),....,v_N(t)$. In the steady state, defined by $\frac{\mathrm{d} V(t)}{\mathrm{d} t}=0$, $\lim_{t\to \infty} v_i(t) = v_{i\infty}$ and $\lim_{t\to \infty} V(t)=V_\infty$, we have 
\begin{equation}
\label{Equ:NIMFA-Steady}
(\tau\text{diag}(1-v_{i\infty})A- I)V_{\infty}=0
\end{equation}
Given the network and the effective infection rate $\tau$, we can numerically compute the infection probability $v_{i\infty}$ as a function of the effective infection rate $\tau$ for each node $i$ by solving (\ref{Equ:NIMFA-Steady}). The trivial, i.e.\ all-zero, solution indicates the absorbing state where all nodes are susceptible. The non-zero solution of $V_\infty$ in (\ref{Equ:NIMFA-Steady}), if exists, points to the existence of a metastable state with a non-zero fraction of infected nodes. Or else, the metastable state can be figured as $0$ or not existing. In this paper, we are interested in actually the metastable state.

The governing equation (\ref{equ_single_NIMFA}) can be extended to the heterogeneous case:
\begin{equation}
\label{equ_heterosingle_NIMFA}
\frac{\mathrm{d} v_i(t)}{\mathrm{d} t}=-\delta v_i(t)+(1-v_i(t))\sum_{j=1}^{N}\beta_{ij}a_{ij}v_{j}(t)
\end{equation}
where $\beta_{ij}=\beta_{ji}$ is the infection rate between node $i$ and $j$. The matrix equation is 
\begin{equation}
\label{Equ:NIMFA-Steady-hetero}
(\frac{1}{\delta}\text{diag}(1-v_{i\infty})BA- I)V_{\infty}=0
\end{equation}
where $B$ is the infection rate matrix with the element $\beta_{ij}$.

\subsection{The i.i.d.\ heterogeneous infection rates}
\label{Chap:Hbeta}
In this paper, we keep the average infection rate to $1$ and tune the recovery rate $\delta$ to control the effective infection rate $\tau$. In the case of the i.i.d.\ heterogeneous infection rates, we aim to explore how the heterogeneous infection rates influence the accuracy of NIMFA when the variance of the infection rate varies. Particularly, we compare the average fraction $y_\infty$ of infected nodes obtained by NIMFA and the simulations for a given effective infection rate $\tau$. In this subsection, we introduce the distribution of the heterogeneous infection rates that will be considered in this work. We choose the infection-rate distribution that is frequently observed in real-world and importantly the variance is tunable with a fixed mean so that we can systematically explore how the accuracy of NIMFA changes with the broadness of the i.i.d.\ infection rate. 

We consider the log-normal distribution, of which we can keep the mean unchanged and tune the variance in a large range. The log-normal distribution \cite{van2006performance} $B\sim Log\textrm{-}\mathcal{N}(\beta;\mu,\sigma)$, of which the probability density function (PDF) is, for $\beta>0$
\begin{equation}
f_B(\beta; \mu, \sigma)=\frac{1}{\beta\sigma \sqrt{2\pi}}exp\left(-\frac{(\ln \beta-\mu)^2}{(2\sigma^2)}\right)
\end{equation}
has a power-law tail for a large range of $\beta$ provided $\sigma$ is sufficiently large. The log-normal distribution has as well been widely observed in real-world, where the interaction frequency between nodes is usually considered as the infection rate between those nodes. For example Wang et al.~\cite{WANGWenBin:2143} find that by employing the log-normal distributed infection rates, their epidemic model can accurately fit the infection data of 2003 SARS; we also find that the infection rates in an airline network follow the log-normal distribution \cite{qu2015sis}.

In \cite{qu2015sis}, we find that, if the epidemic does not die out, the larger the variance of the i.i.d.\ infection rate is, the smaller the average fraction $y_\infty$ of infected nodes is. We will show that this conclusion can actually explain the observation about how the accuracy of NIMFA changes with the variance of the i.i.d.\ infection rates at a given effective infection rate $\tau$ in this paper.  

\subsection{The correlated heterogeneous infection rates and the range of $\alpha$}
\label{sec_ctau}
In the case of the correlated heterogeneous infection rates, we build a correlated infection-rate scenario and a reference scenario. In the scenario of correlated infection rates, we assume that $\beta _{ij}=c\left( d_{i}d_{j}\right) ^{\alpha }$ where $c$ is selected so that the average infection rate is $1$ and $\alpha$ indicates the correlation strength. In this case, the infection rate of each link is determined by the given network topology and $\alpha$. For the reference scenario, we shuffle the infection rates from all the links as generated in the first scenario and redistribute them randomly to all the links. In this way, we keep the distribution of infection rates but effectively remove the correlation between the infection rates and nodal degrees. For simplicity, we name this reference scenario as the uncorrelated infection-rate scenario. Though the i.i.d.\ infection rates are also uncorrelated, we can tune the variance of the infection rate in the case of the i.i.d.\ infection rates while keeping the distribution and the mean of the infection rates. However, in the scenario of uncorrelated infection rates in this paper, the distribution of the infection rate changes with the parameter $\alpha$, hence the variance of the heterogeneous infection rates cannot be systematically tuned. 

A positive $\alpha>0$ (or negative $\alpha<0$), suggests a positive (or negative) correlation between the infection rates and nodal degrees. Too large or too small values of $\alpha$ could not be realistic. For example, \cite{barrat2004architecture,macdonald2005minimum,li2004statistical} suggest that $\alpha$ is around $0.5$ or $0.8$ in their datasets. Hence, we select $\alpha=-0.25,-0.5,-1$ for the negative correlation and $\alpha=0.25,0.5,1$ for the positive correlation. Different values of $\alpha$ also offer the possibility to explore how NIMFA performs when the correlation strength is different.

In \cite{qu2016heterogeneous}, we find that, comparing to the scenario of uncorrelated infection rate, 1) the positive correlation between the infection rate and the nodal degree tends to increase (or decrease) the average fraction of infected nodes when the effective infection rate $\tau$ is small (or large); 2) the negative correlation tends to decrease (or increase) the average fraction $y_\infty$ of infected nodes when $\tau$ is small (or large). In this paper, we aim to understand how the correlation influences the accuracy of NIMFA by comparing the average fraction $y_\infty$ of infected nodes obtained by NIMFA and the simulations of the exact SIS model. As in the case of the i.i.d.\ infection rates, we will show that the influence of the correlation between the infection rate and the nodal degree on the average fraction $y_\infty$ of infected nodes can also be used to partially explain the conclusions in this paper. 

\subsection{The simulations}
We perform the continuous-time simulations of the SIS model on both ER networks and SF networks in this paper. 
We develop the continuous-time simulator for the SIS model with heterogeneous infection rates, based on the one firstly proposed by van de Bovenkamp and described in detail in \cite{li2012susceptible} for homogeneous infection rates.
Given a network topology, a recovery rate $\delta$, we carry out $100$ iterations.
In each iteration, we construct the network as described in Section \ref{sec_top}.
We generate the i.i.d.\ heterogeneous infection rates following the log-normal distribution or the correlated heterogeneous infection rates as described in (\ref{eq1}) for the scenario of the correlated infection rates and shuffle them for the scenario of uncorrelated infection rates. Initially, $10\%$ of the nodes are randomly infected.
Then the infection and recovery processes of SIS model are simulated until the system reaches the metastable state where the fraction of infected nodes is nonzero and unchanged for a long time if the epidemic spreads out, or the fraction is zero if the epidemic dies out.
The average fraction $y_\infty$ of infected nodes is obtained over $100$ iterations. 

\section{Effect of the heterogeneous infection rates}
In this section, we first explore the accuracy of NIMFA when the heterogeneous infection rates are i.i.d., and particularly how NIMFA performs when the variance Var$[B]$ of the infection rate $B$ varies.
Then we explore the influence of the correlated infection rates on NIMFA.

\subsection{The i.i.d.\ infection rates}
\begin{figure}[!t]
\centering
\subfigure[]{
\includegraphics[scale=.28]{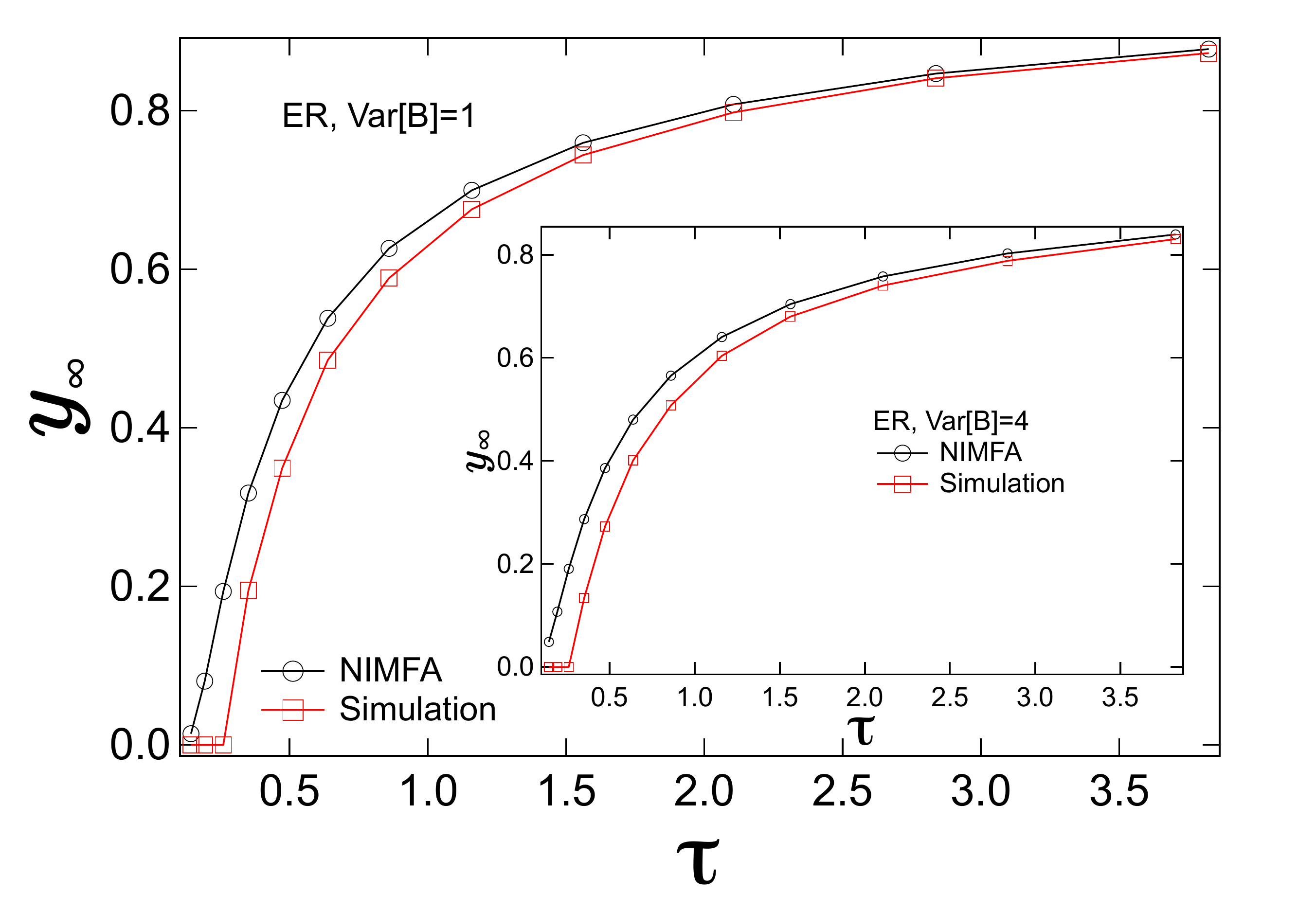}
\label{fig:11}
}
\subfigure[]{
\includegraphics[scale=.28]{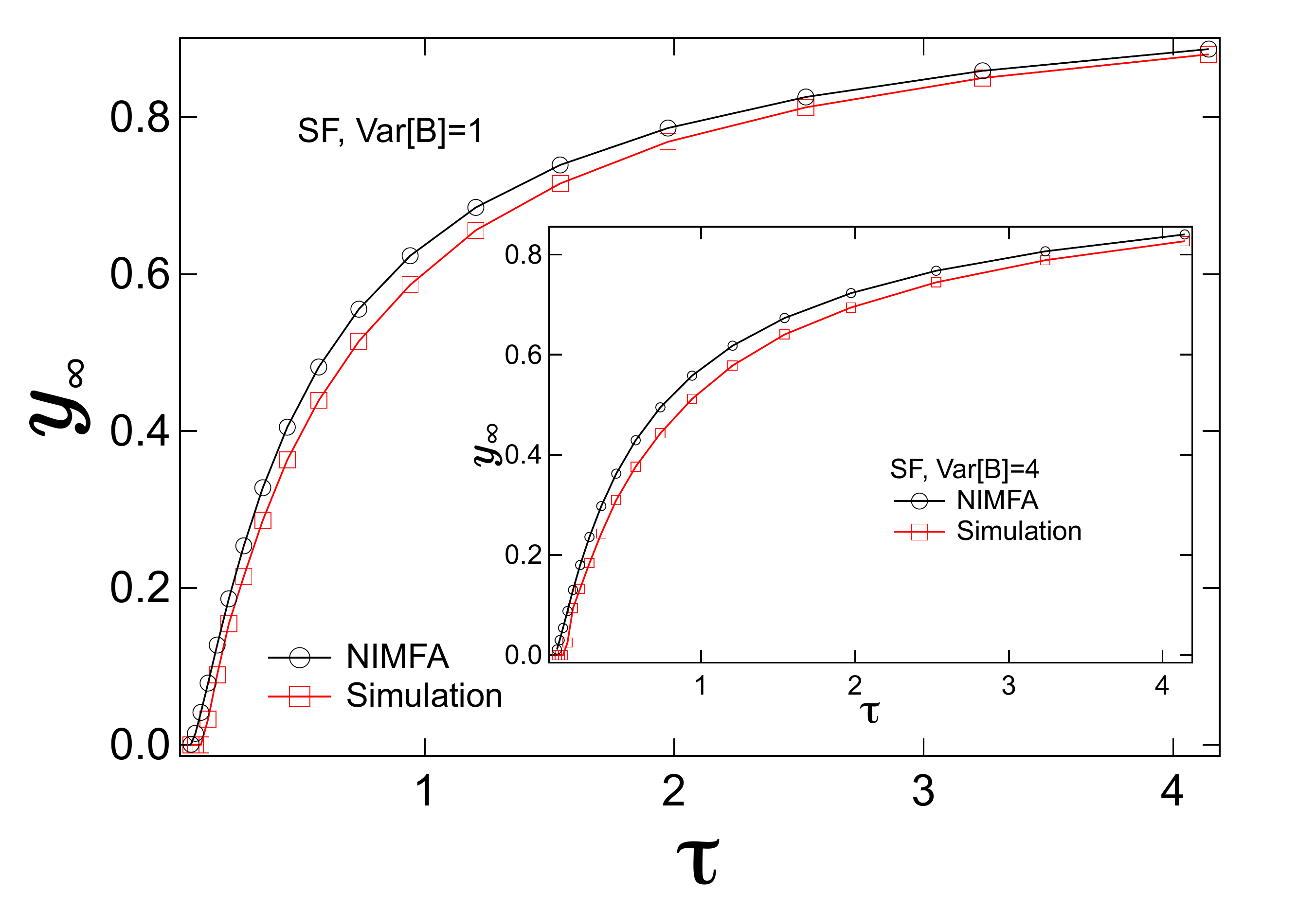}
\label{fig:12}
}
\caption{The average fraction $y_\infty$ as a function of the effective infection rate $\tau$ for (a) ER networks and (b) SF networks. The variances of the infection rates are $1$ and $4$ in the main figure and the inset respectively.}
\label{fig:10}
\end{figure}

We aim to understand the precision of NIMFA under different effective infection rates, different variances of infection rates and different network topologies: we set the average infection rate to $1$ and tune the recovery rate $\delta$ to control the effective infection rate $\tau$; we change the variance of infection rates which follow the log-normal distribution; we consider both ER and SF networks to represent different topologies.
For each value of the variance of the infection rate, we obtain the average fraction $y_\infty$ of infected nodes as a function of the effective infection rate $\tau$ for NIMFA by numerically solving (\ref{Equ:NIMFA-Steady-hetero}) and compare with that by the continuous-time simulations. 
As shown in Fig.~\ref{fig:10}, no matter what the variance of the infection rate is, the curve of $y_\infty$ vs.~$\tau$ obtained by NIMFA is close to that obtained by simulations when the actual prevalence of the epidemic is high, i.e.\ the effective infection rate $\tau$ is large.

In order to quantify the difference between the two curves obtained by NIMFA and simulations, we define the variable $\zeta$:
\begin{equation}
\label{equ_zeta}
\zeta(\tau)=\frac{\lvert y_{\infty,N}(\tau)-y_{\infty,S}(\tau)\rvert}{y_{\infty,S}(\tau)}
\end{equation}
where $y_{\infty,N}(\tau)>0$ and $y_{\infty,S}(\tau)> 0$ denote the average fraction of infected nodes obtained by NIMFA and simulations respectively. The larger the value of $\zeta(\tau)$ is, the less accurate NIMFA is at the corresponding $\tau$. 

\begin{figure}[!t]
\centering
\subfigure[]{
\includegraphics[scale=.28]{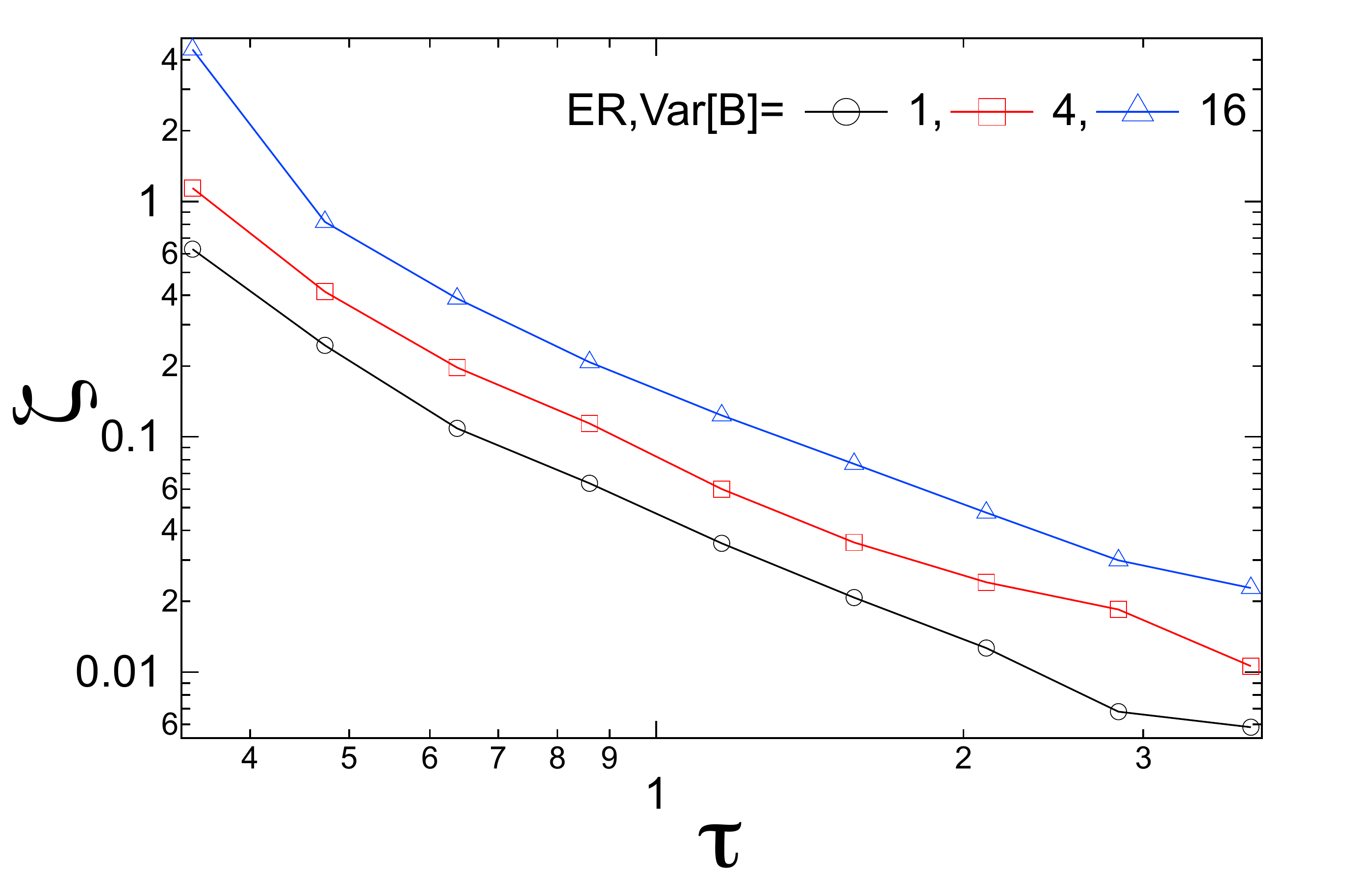}
\label{fig:21}
}
\subfigure[]{
\includegraphics[scale=.28]{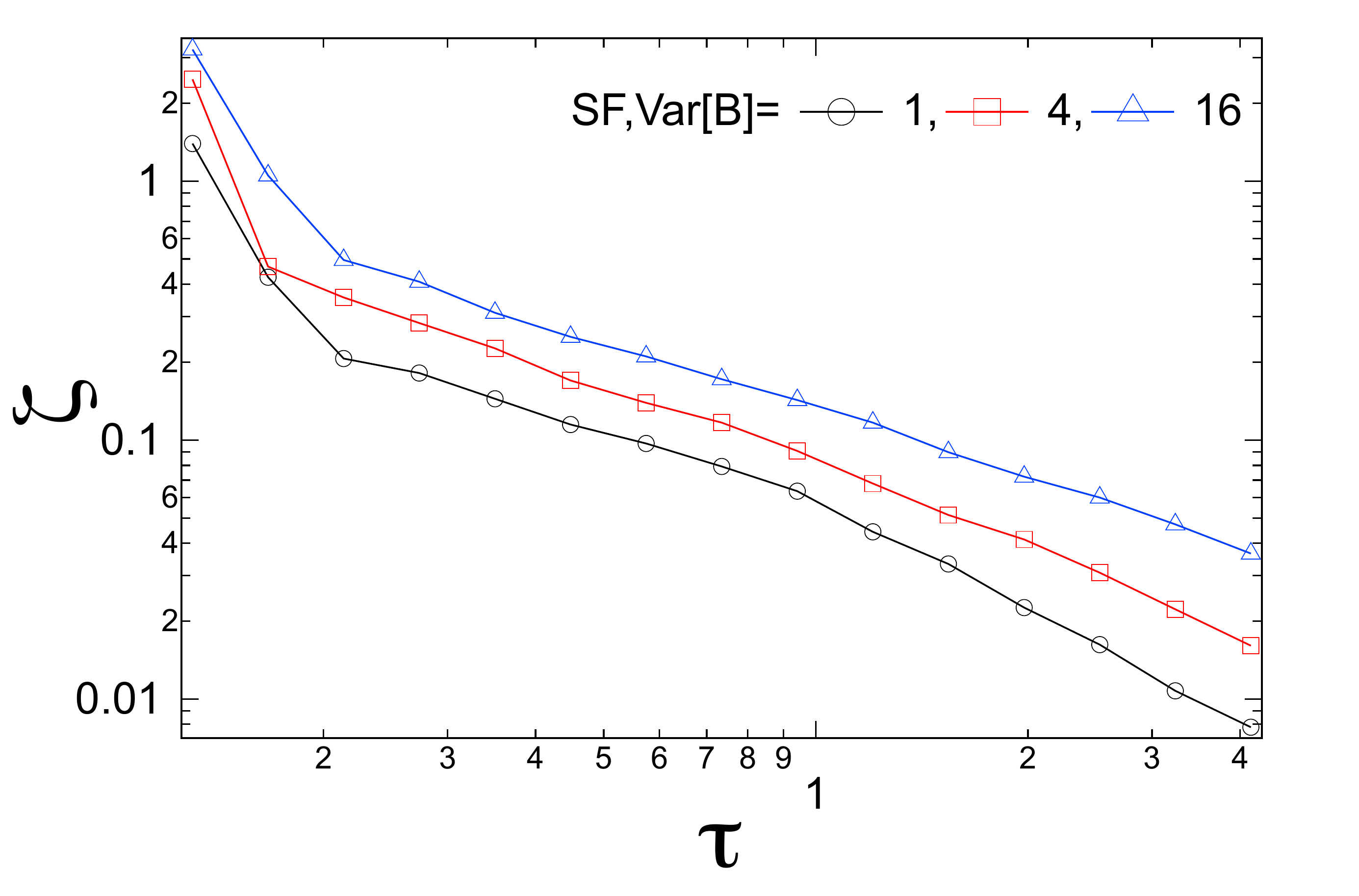}
\label{fig:22}
}
\caption{The plot of $\zeta$ as a function of the effective infection rate $\tau$ for (a) ER networks and (b) SF networks.}
\label{fig:20}
\end{figure}

In Fig.~\ref{fig:20}, the plots of $\zeta$ vs.~$\tau$ are shown for both ER and SF networks. We find that, for a given effective infection rate $\tau$, NIMFA becomes less accurate when the variance of the i.i.d.\ heterogeneous infection rates increases. This observation can be to a large extent explained by: 1) our finding in Fig.~\ref{fig:10} that NIMFA is more accurate when the prevalence is higher; 2) that given an effective infection rate $\tau$ a smaller variance of the i.i.d.\ infection rates leads to a higher prevalence \cite{qu2015sis}.

\begin{figure}[!t]
\centering
\subfigure[]{
\includegraphics[scale=.28]{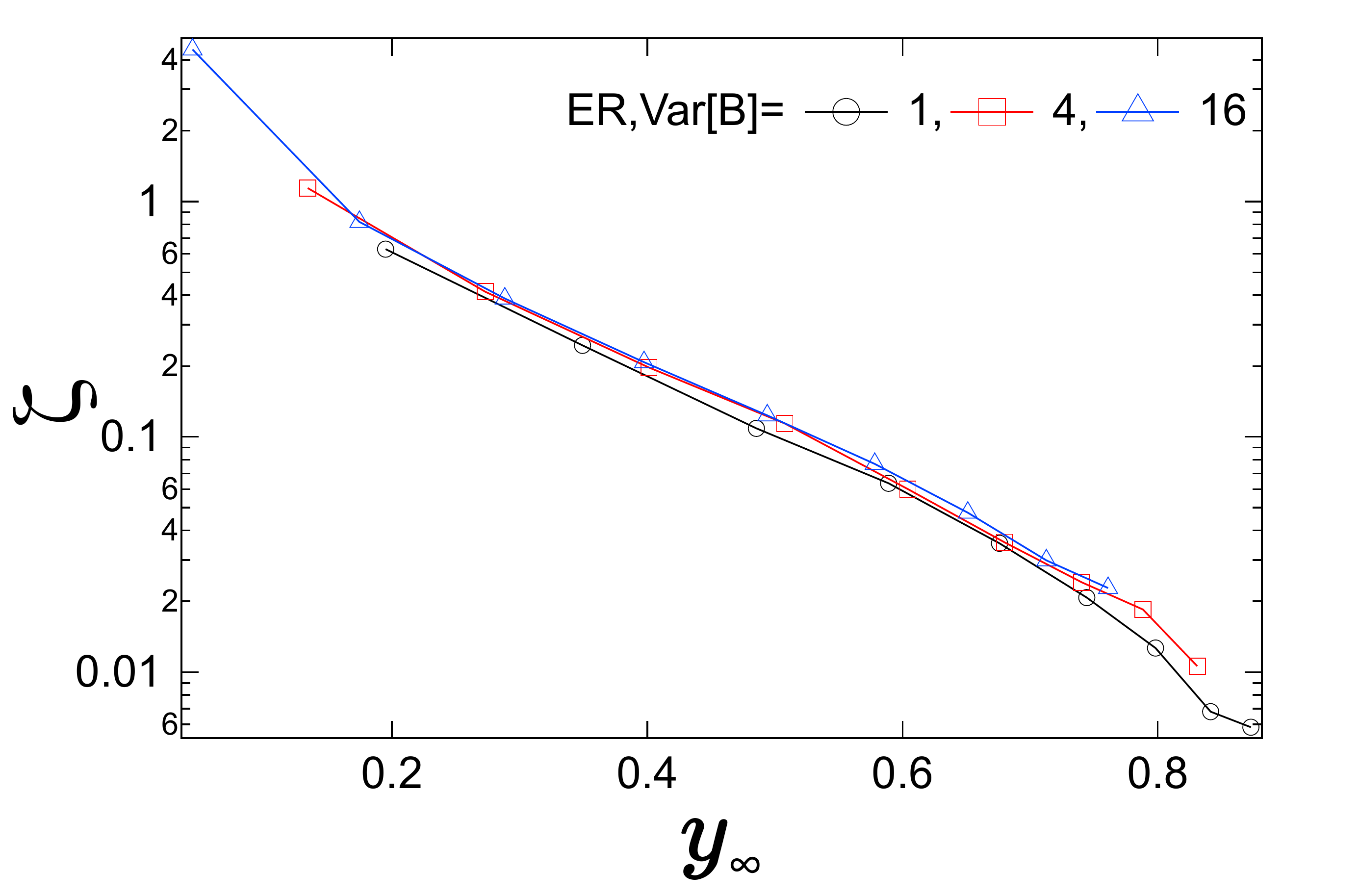}
\label{fig:31}
}
\subfigure[]{
\includegraphics[scale=.28]{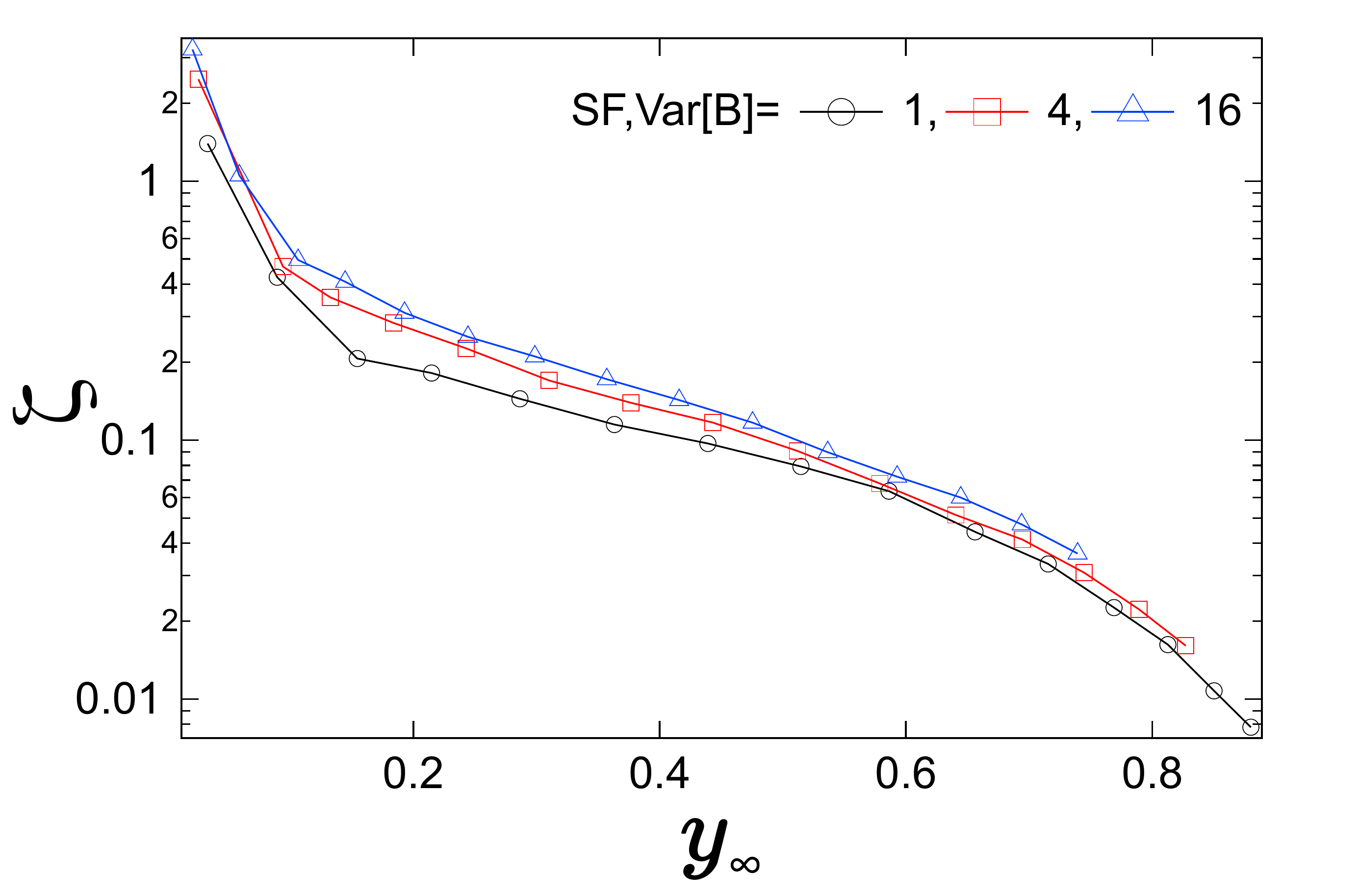}
\label{fig:32}
}
\caption{The plot of $\zeta$ as a function of the average fraction $y_\infty$ obtained by simulations for (a) ER networks and (b) SF networks.}
\label{fig:30}
\end{figure}

We further explore how the variance of the infection rates influences the accuracy of NIMFA if the actual prevalence $y_{\infty,S}(\tau)$ of epidemic is similar.
We plot the variable $\zeta$ in (\ref{equ_zeta}) as a function of the actual average fraction of infected nodes obtained by simulations in Fig.~\ref{fig:30}.
We find that though it is less evident for ER networks in Fig.~\ref{fig:31}, the difference $\zeta$ in (\ref{equ_zeta}) is actually larger if the variance of the infection rate is larger as shown in Fig.~\ref{fig:32} for SF networks when the prevalence is the same. Hence, the higher heterogeneity,  i.e.\ the larger variance, of the i.i.d.\ infection rates tends to lower down more the accuracy of NIMFA.
Overall, we conclude that the prevalence of the epidemic mainly affects the accuracy of NIMFA, i.e.\ the higher the prevalence is, the more accurate NIMFA tends to be, and given the same prevalence, a larger variance of the i.i.d.\ infection rates tends to lower down the accuracy of NIMFA. 

\subsection{The correlated infection rate}
\label{sec:3.1} 
In this subsection, we aim to understand how the correlation between the infection rate and the nodal degree as shown in (\ref{eq1}) influences the accuracy of NIMFA. We first employ ER networks as an example and discuss the case when the correlation is positive. Afterwards we explore the influence of the negative correlation. 

\begin{figure}
\centering
\subfigure[]{
\includegraphics[scale=.28]{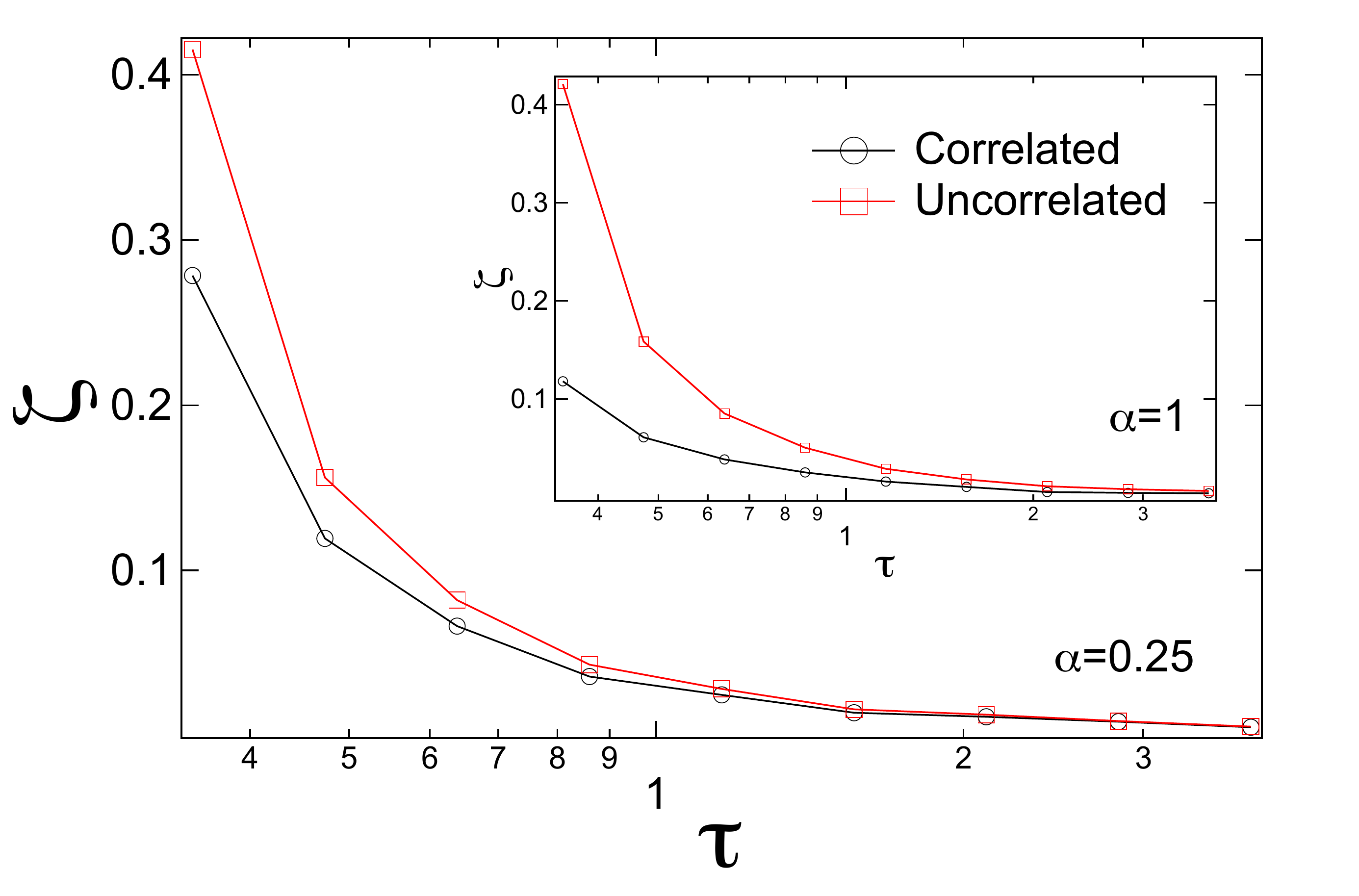}
\label{fig:41}
}
\subfigure[]{
\includegraphics[scale=.28]{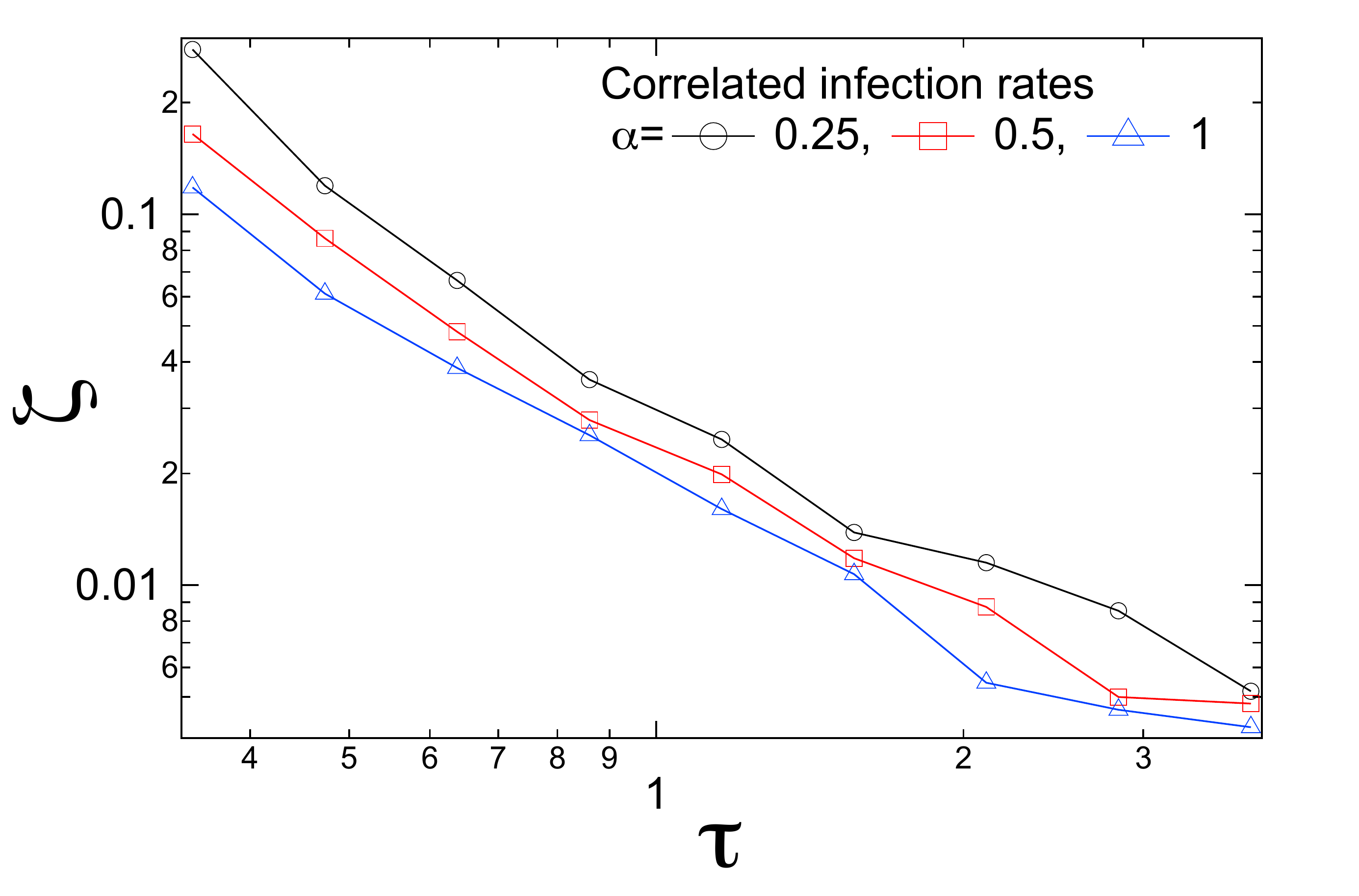}
\label{fig:42}
}
\subfigure[]{
\includegraphics[scale=.28]{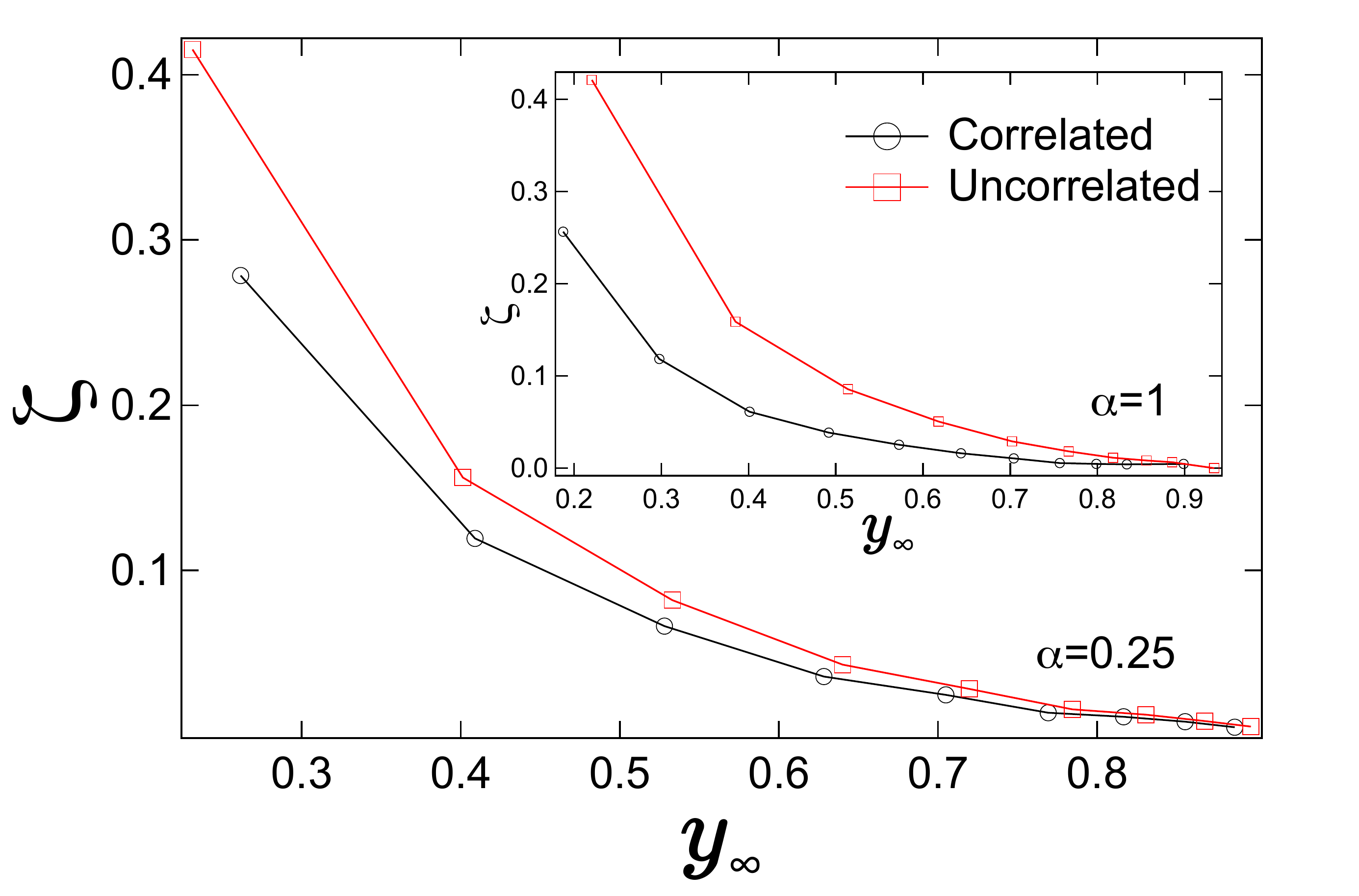}
\label{fig:43}
}
\subfigure[]{
\includegraphics[scale=.28]{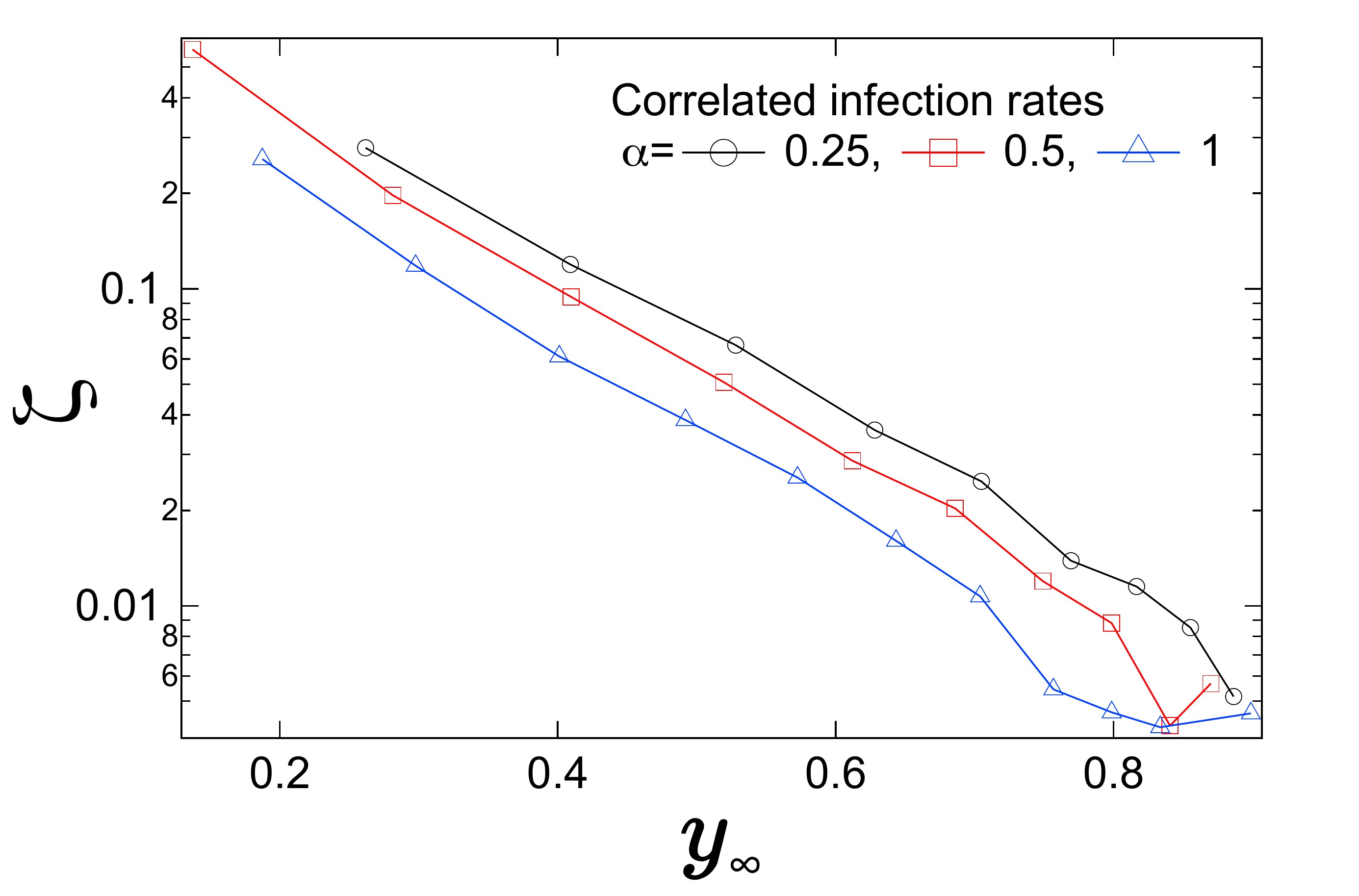}
\label{fig:44}
}
\caption{(a) The plot of $\zeta$ as a function of the effective infection rate $\tau$ in the scenarios of uncorrelated and correlated infection rates for $\alpha=0.25$ (the main figure) and $\alpha=1$ (the inset). (b) The plot of $\zeta$ as a function of the effective infection rate $\tau$ in the scenario of correlated infection rates where different values of $\alpha$ are considered. (c) The plot of $\zeta$ as a function of the average fraction $y_\infty$ of infected nodes obtained by simulations in the scenarios of uncorrelated and correlated infection rates for $\alpha=0.25$ (the main figure) and $\alpha=1$ (the inset). (d) The plot of $\zeta$ as a function of the effective infection rate in the scenario of correlated infection rates where different values of $\alpha$ are considered. All the plots are on ER networks.}
\label{fig:40}
\end{figure}

As mentioned in Section \ref{sec_ctau}, we build the scenario of uncorrelated infection rates as a reference to study the influence of the correlation between the infection rate and the nodal degree by shuffling the infection rates from all the links as generated in the scenario of correlated infection rates and redistributing them randomly to all the links. 
As shown in Fig.~\ref{fig:41}, we compare the difference $\zeta$ between NIMFA and simulations in the scenario of uncorrelated and correlated infection rates for both $\alpha=0.25$ and $\alpha=1$, and find that $\zeta$ is smaller in the scenario of correlated infection rates, i.e.\ NIMFA is more accurate at a given a given effective infection rate $\tau$ when the correlation between the infection rate and the nodal degree is positive comparing to the scenario of uncorrelated infection rates.
The observations are also consistent with our conclusion that NIMFA is more accurate when the prevalence is higher: the positive correlation tends to increase the average fraction of infected nodes \cite{qu2016heterogeneous}, and thus the accuracy of NIMFA, when the effective infection rate $\tau$ is small; however, when the effective infection rate $\tau$ is large, though the positive correlate may lower down a bit the average fraction $y_\infty$ of infected nodes, the prevalence in both scenarios is high, i.e.\ NIMFA is relatively accurate, and the difference of the accuracy of NIMFA in the two scenarios is not obvious.   
As the correlation strength $\alpha$ increases in Fig.~\ref{fig:42}, the difference $\zeta$ decreases at a given $\tau$. That is to say, NIMFA tends to be more accurate when the positive correlation becomes stronger. 

We further consider the influence of the positive correlation on the accuracy of NIMFA when the prevalence is the same. The plots of the difference $\zeta$ as a function of the average fraction $y_\infty$ of infected nodes are shown in Fig.~\ref{fig:43} and Fig.~\ref{fig:44}.
Given the prevalence of epidemic, the positive correlation is more likely to increase the precision of NIMFA and the stronger the correlation is the more accurate NIMFA is. 
We observe the same on SF networks which is though not shown here. 

\begin{figure}
\centering
\subfigure[]{
\includegraphics[scale=.28]{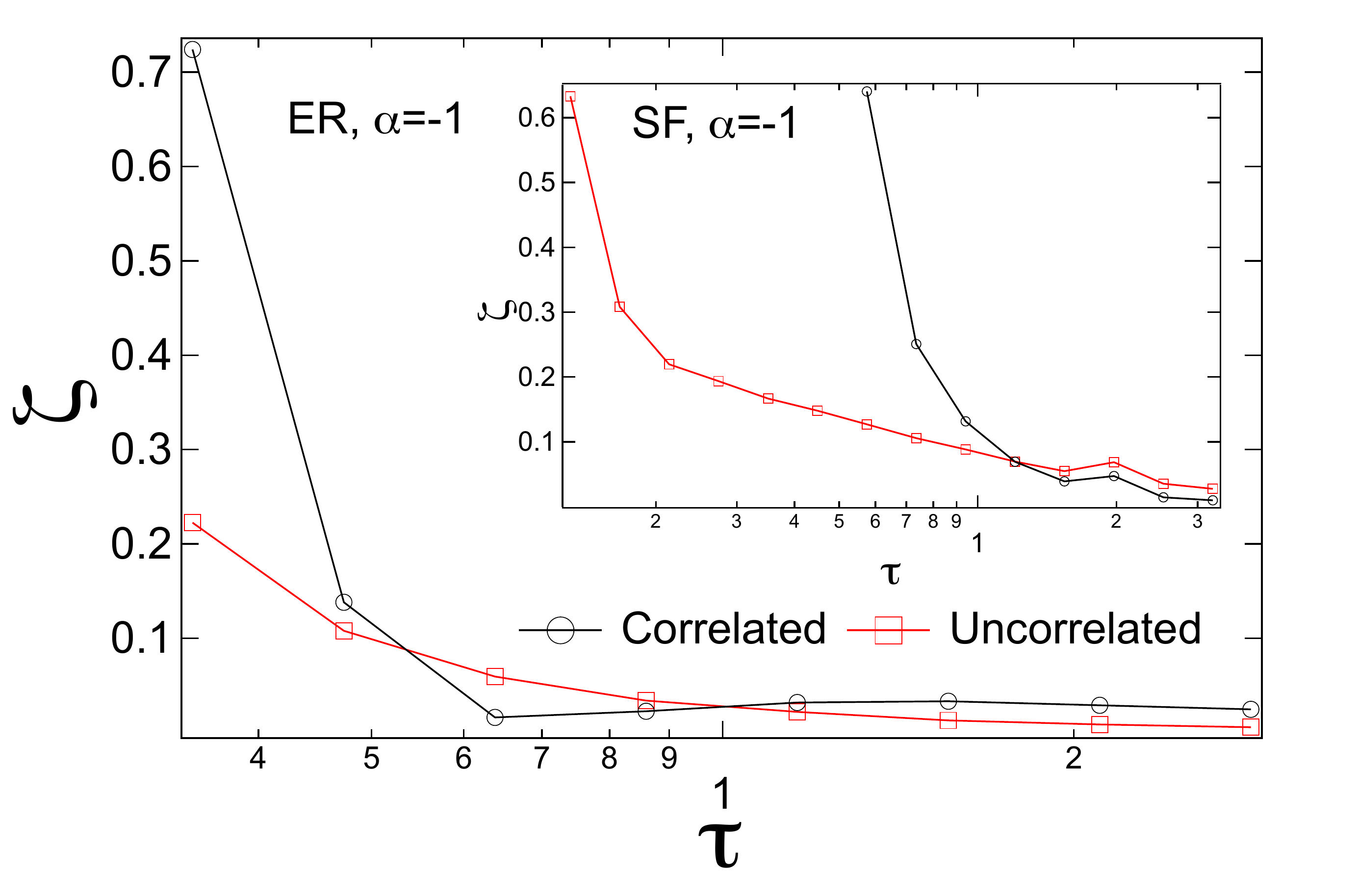}
\label{fig:51}
}
\subfigure[]{
\includegraphics[scale=.28]{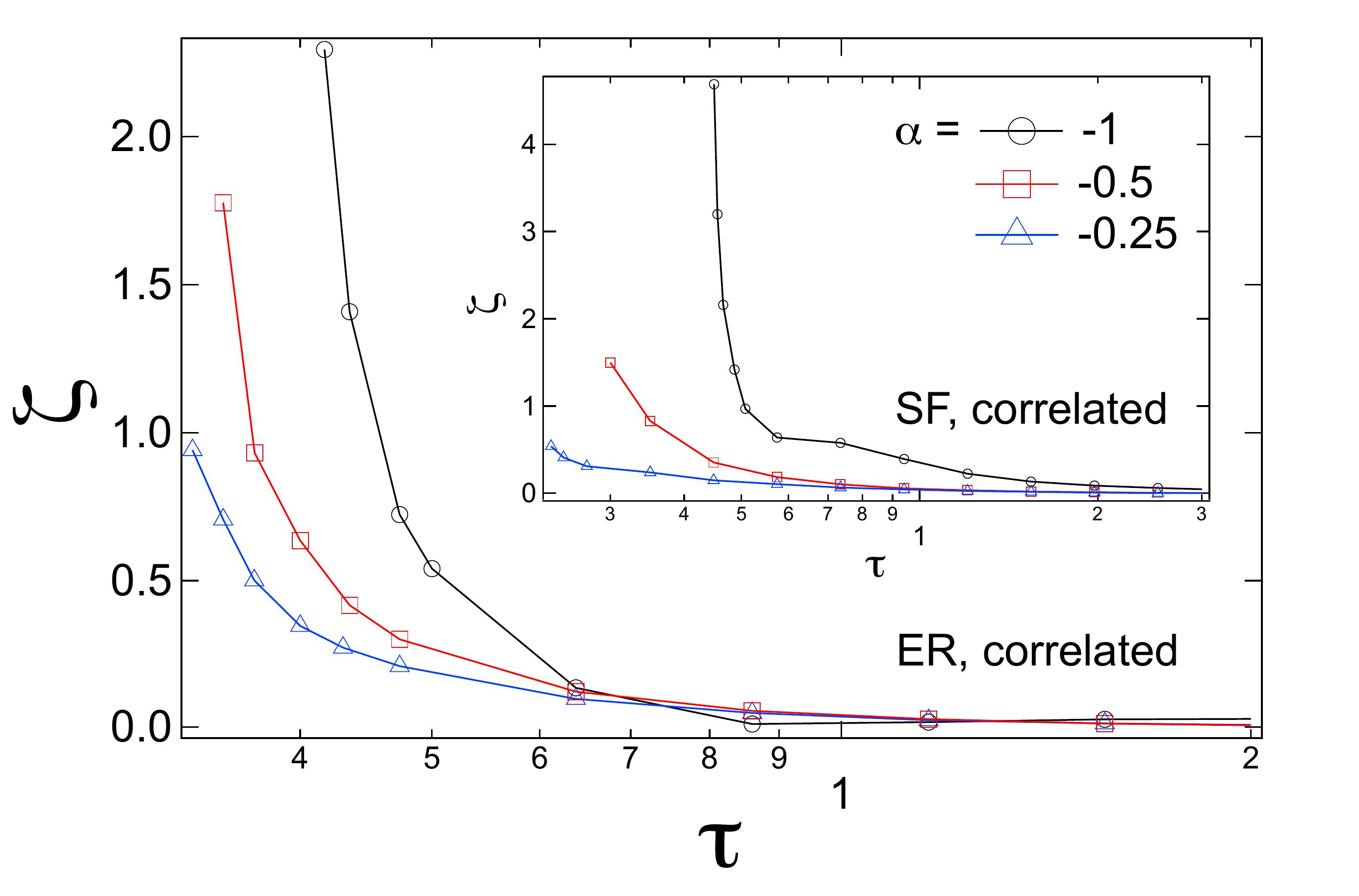}
\label{fig:52}
}
\subfigure[]{
\includegraphics[scale=.28]{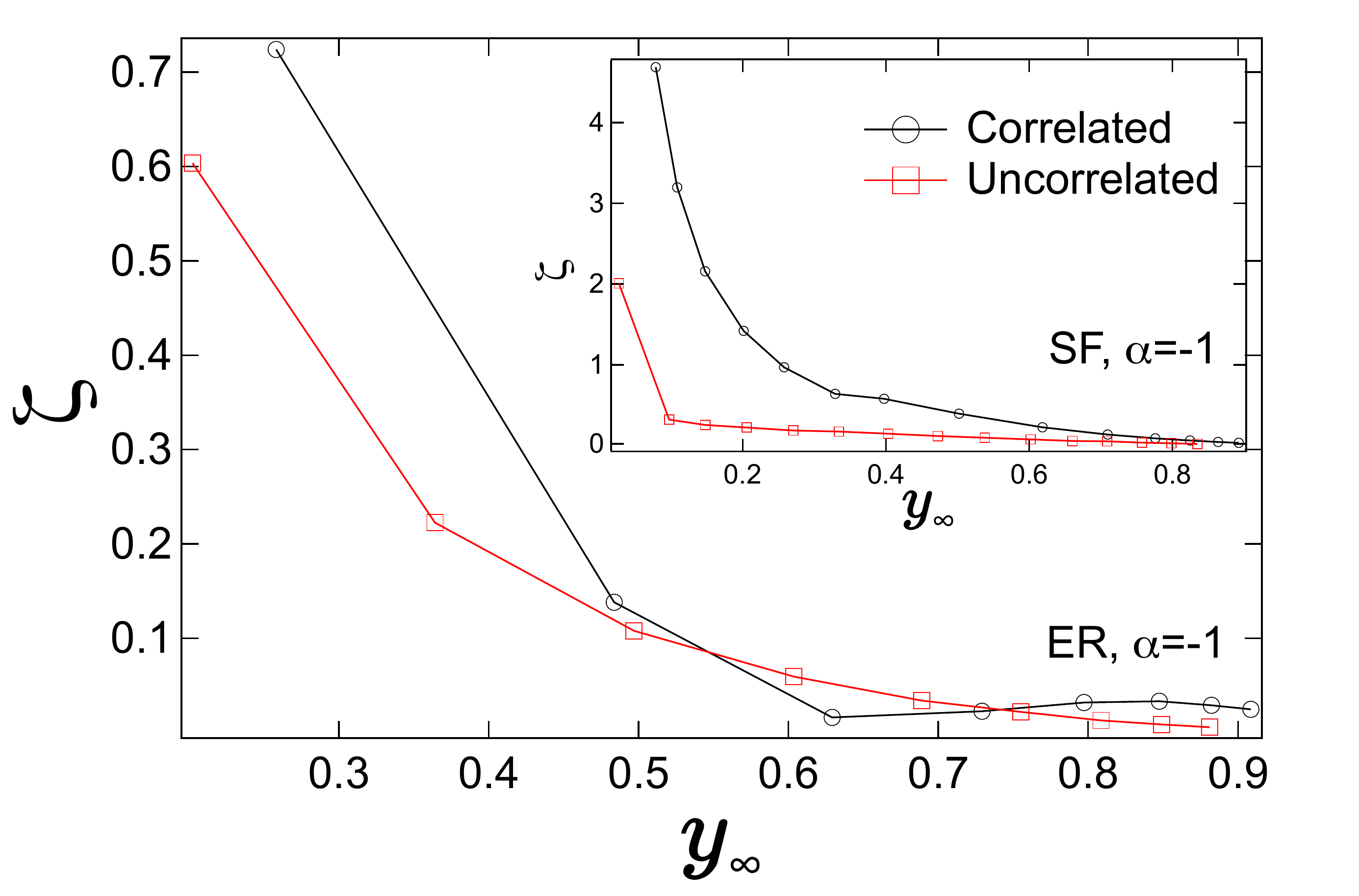}
\label{fig:53}
}
\subfigure[]{
\includegraphics[scale=.28]{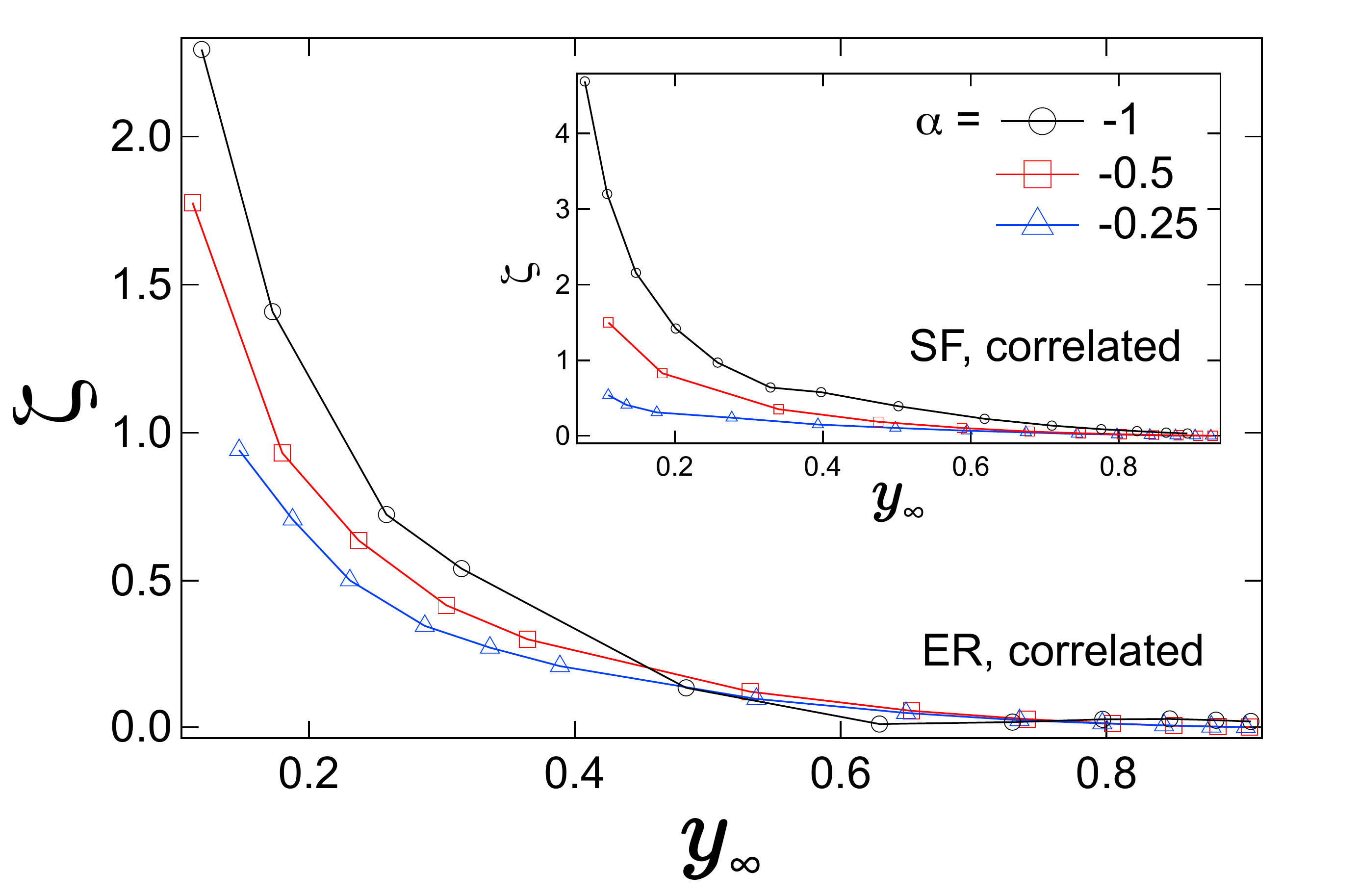}
\label{fig:54}
}
\caption{(a) The plot of $\zeta$ as a function of the effective infection rate $\tau$ in the scenarios of uncorrelated and correlated infection rates for $\alpha=-1$. (b) The plot of $\zeta$ as a function of the effective infection rate $\tau$ in the scenario of correlated infection rates where different values of $\alpha$ are considered. (c) The plot of $\zeta$ as a function of the average fraction $y_\infty$ of infected nodes obtained by simulations in the scenarios of uncorrelated and correlated infection rates for $\alpha=-1$. (d) The plot of $\zeta$ as a function of the average fraction $y_\infty$ of infected nodes obtained by simulations in the scenario of correlated infection rates where different values of $\alpha$ are considered.}
\label{fig:50}
\end{figure}

Regarding to the influence of the negative correlation between the infection rate and the nodal degree on the accuracy of NIMFA, we compare the variable $\zeta$ in the scenario of correlated and uncorrelated infection-rate scenario with $\alpha=-1$ for both ER and SF networks as shown in Fig.~\ref{fig:51}. We find that, in general, the negative correlation significantly decreases the accuracy of NIMFA when the effective infection rate $\tau$ is small but may slightly increase that when $\tau$ is large. Moreover, NIMFA becomes less accurate when the negative correlation is stronger as shown in Fig.~\ref{fig:52}. 
As mentioned in Section \ref{sec_ctau}, the negative correlation tends to decrease the prevalence when the effective infection rate $\tau$ is small while increase the prevalence when $\tau$ is large. Hence, the influence of prevalence on the precision of NIMFA could largely explain our observations here. 

When the prevalence of epidemic is the same, the influence of the negative correlation on NIMFA's accuracy is shown in Fig.~\ref{fig:53} and Fig.~\ref{fig:54}. We find that, in general, 1) NIMFA is less accurate with the negative correlation comparing to the uncorrelated scenario especially when the prevalence is low as shown in Fig.~\ref{fig:53}; 2) NIMFA becomes even less accurate if the negative correlation becomes stronger as shown in Fig.~\ref{fig:54}. 

\section{Real-world network}
The interaction frequency between two nodes in a real-world network has been considered as the infection rate between the pair of nodes \cite{qu2015sis}. In this section, we choose the airline network from the real world as an example to illustrate how its heterogeneous infection rates affect the accuracy of NIMFA of SIS epidemics on the network. 

In the airline network, the nodes are the airports, the link between two nodes indicates that there's at least one flight between these two airports, and the infection rate along a link is the number of flights between the two airports. We construct this network and its infection rates from the dataset of openFlights\footnote{http://openflights.org/data.html}. As shown in \cite{qu2016heterogeneous}, the airline network possess roughly a power-law degree distribution.
The heterogeneous infection rates from the dataset are normalized by the average so that the average is $1$.
We compare the difference $\zeta$ between NIMFA and the simulations of the exact SIS model in three scenarios: 1) the network is equipped with its normalized original heterogeneous infection rates (correlated) as given in the dataset; 2) the network is equipped with the infection rates in the normalized original dataset but randomly shuffled (uncorrelated); 3) the network is equipped with a constant infection rate (homogeneous) which equals to $1$. The original heterogeneous infection rate between a pair of nodes are approximately correlated with the degrees of the two nodes as the relationship (\ref{eq1}), and the parameter $\alpha\approx 0.14$ indicates a positive correlation \cite{qu2016heterogeneous}.

\begin{figure}
\centering
\subfigure[]{
\includegraphics[scale=.28]{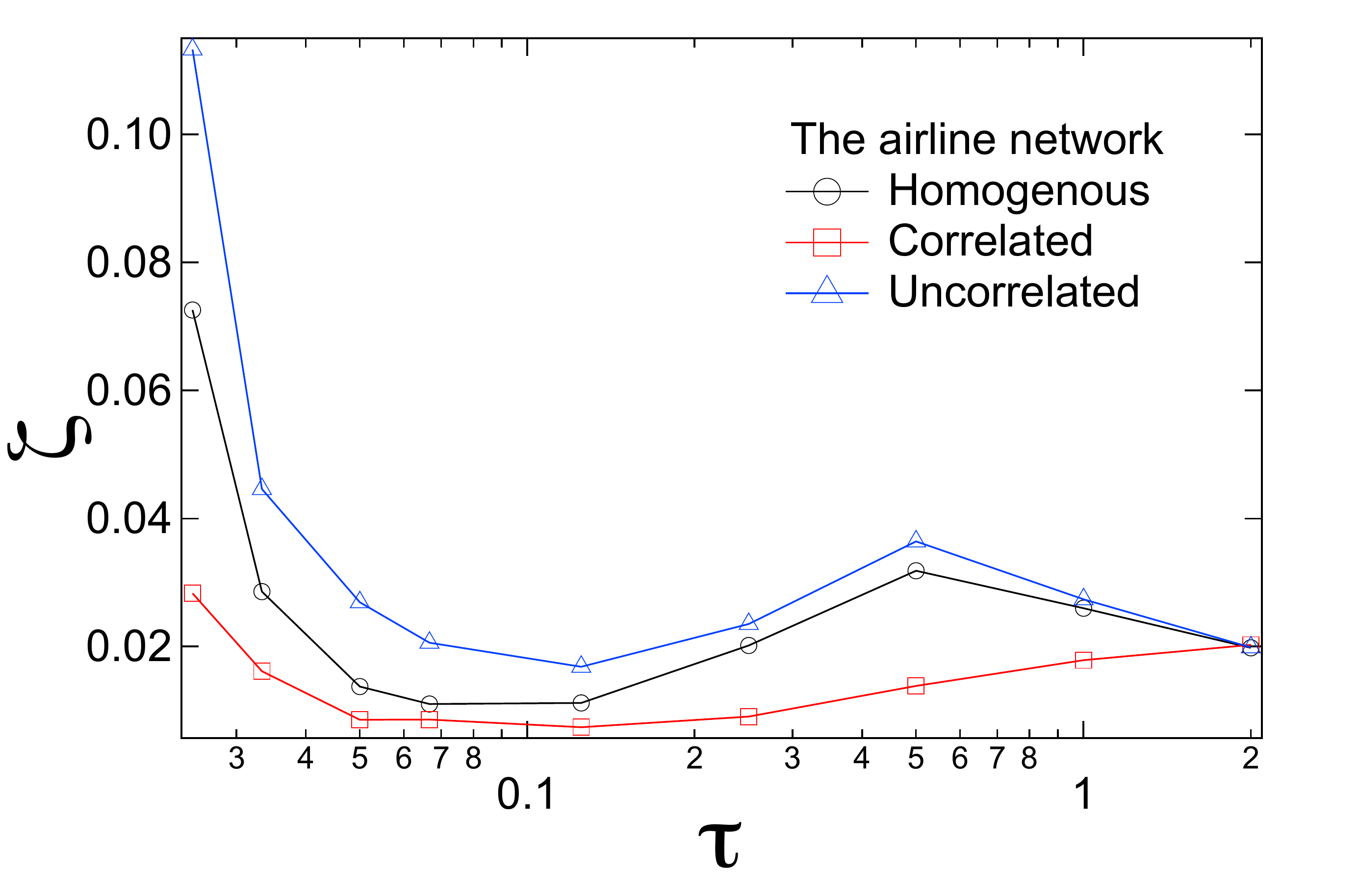}
\label{fig:71}
}
\subfigure[]{
\includegraphics[scale=.28]{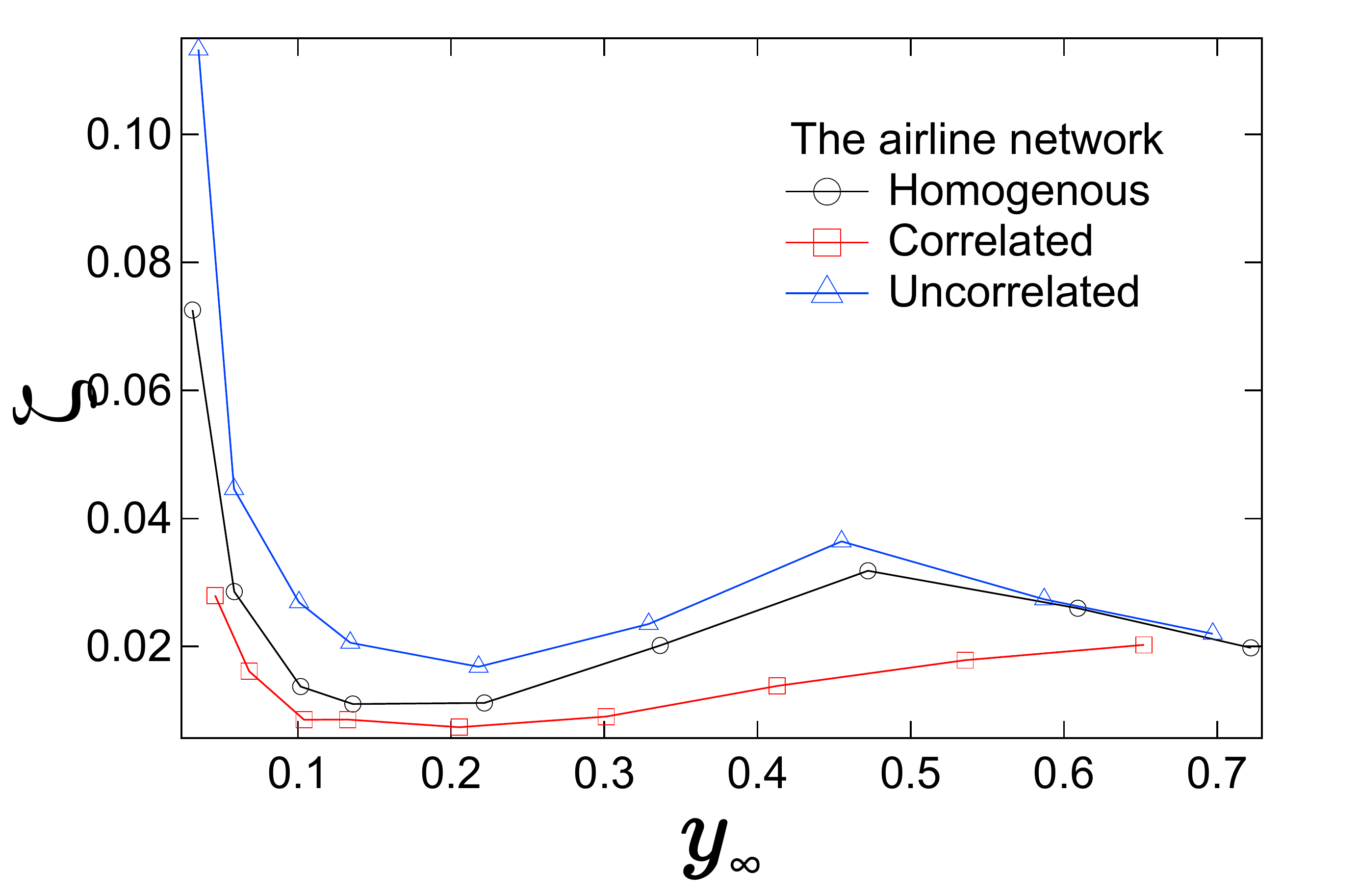}
\label{fig:72}
}
\caption{The plot of $\zeta$ as a function of (a) the effective infection rate $\tau$ and (b) the average fraction $y_\infty$ of infected nodes obtained by simulations in the airline network with different scenarios of infection rates.}
\label{fig:70}
\end{figure}

We show the difference $\zeta$ as a function of the effective infection rate $\tau$ in Fig.~\ref{fig:71} for the 3 scenarios defined as above.
We find that NIMFA is generally more accurate when the effective infection rate $\tau$ is larger, i.e.\ the prevalence of epidemic is high.
The variable $\zeta$ is smaller in the scenario of homogeneous infection rates than uncorrelated infection rates with any effective infection rate. 
This is because the i.i.d.\ infection rates with a non-zero variance tends to decrease the prevalence, and thus lower down the accuracy of NIMFA at a given effective infection rate $\tau$. 
NIMFA is more accurate with the positive correlation by comparing the difference $\zeta$ in the scenario of correlated infection rates and uncorrelated infection rates.
Furthermore, Fig.~\ref{fig:72} shows that, given the same actual prevalence, i.e.\ the average fraction $y_\infty$ of infected nodes obtained by simulations, NIFMA is more accurate: 1) in the homogeneous scenario than in the uncorrelated scenario; 2) in the correlated scenario than in the uncorrelated scenario. 
All the observations agree with our previous observations and explanations about how the heterogeneous infection rate influences the accuracy of NIMFA in network models.    

\section{Conclusion}
In this paper, we study how the heterogeneous infection rates affect the accuracy of NIMFA -- an advanced mean-field approximation of SIS model that takes the underly network topology into account. By comparing NIMFA with the continuous-time simulations of the exact SIS model at a give effective infection rate $\tau$, we find that the prevalence of epidemic could largely characterize the accuracy of NIMFA which is reflected in two aspects: 1) NIFMA is generally more accurate when the $\tau$ is larger, i.e.\ the prevalence of epidemic is higher; 2) when the variance of the i.i.d.\ infection rates or the correlation between the infection rate and the nodal degree decreases the prevalence at a given $\tau$, NIMFA tends to become less accurate as well.
Moreover, we also explore the influence of the heterogeneous infection rates on the accuracy of NIMFA at a given prevalence, i.e.\ when the average fraction $y_\infty$ of infected nodes obtained by simulations is given. Regarding to the i.i.d.\ heterogeneous infection rates, the accuracy of NIMFA tends to decrease as the variance of infection rates increases. In the scenario of correlated infection rates, the positive correlation between the nodal degree and the infection rate is more likely to increase the accuracy of NIMFA whereas the negative correlation tends to lower down the accuracy especially when the effective infection rate$\tau$ is small. Our work sheds light on the conditions when we the mean-field approximation of the SIS model with heterogeneous infection rates is accurate.  




\begin{thebibliography}{10}

\bibitem{daley2001epidemic}
Daryl~J Daley, Joe Gani, and Joseph~Mark Gani.
\newblock {\em Epidemic modelling: an introduction}, volume~15.
\newblock Cambridge University Press, 2001.

\bibitem{pastor2001epidemic}
Romualdo Pastor-Satorras and Alessandro Vespignani.
\newblock Epidemic dynamics and endemic states in complex networks.
\newblock {\em Physical Review E}, 63(6):066117, 2001.

\bibitem{wang2013effect}
Huijuan Wang, Qian Li, Gregorio D’Agostino, Shlomo Havlin, H~Eugene Stanley,
  and Piet Van~Mieghem.
\newblock Effect of the interconnected network structure on the epidemic
  threshold.
\newblock {\em Physical Review E}, 88(2):022801, 2013.

\bibitem{0295-5075-105-6-68004}
Daqing Li, Pengju Qin, Huijuan Wang, Chaoran Liu, and Yinan Jiang.
\newblock Epidemics on interconnected lattices.
\newblock {\em EPL (Europhysics Letters)}, 105(6):68004, 2014.

\bibitem{liu2015epidemics}
Meng Liu, Daqing Li, Pengju Qin, Chaoran Liu, Huijuan Wang, and Feilong Wang.
\newblock Epidemics in interconnected small-world networks.
\newblock {\em PloS one}, 10(3):e0120701, 2015.

\bibitem{pastor2014epidemic}
Romualdo Pastor-Satorras, Claudio Castellano, Piet Van~Mieghem, and Alessandro
  Vespignani.
\newblock Epidemic processes in complex networks.
\newblock {\em arXiv preprint arXiv:1408.2701}, 2014.

\bibitem{Pastor-Satorras2001}
Romualdo Pastor-Satorras and Alessandro Vespignani.
\newblock Epidemic spreading in scale-free networks.
\newblock {\em Physical Review Letters}, 86(14):3200, 2001.

\bibitem{van2009virus}
Piet Van~Mieghem, Jasmina Omic, and Robert Kooij.
\newblock Virus spread in networks.
\newblock {\em IEEE/ACM Transactions on Networking}, 17(1):1--14, 2009.

\bibitem{eames2002modeling}
Ken~TD Eames and Matt~J Keeling.
\newblock Modeling dynamic and network heterogeneities in the spread of
  sexually transmitted diseases.
\newblock {\em Proceedings of the National Academy of Sciences},
  99(20):13330--13335, 2002.

\bibitem{gleeson2011high}
James~P Gleeson.
\newblock High-accuracy approximation of binary-state dynamics on networks.
\newblock {\em Physical Review Letters}, 107(6):068701, 2011.

\bibitem{boguna2013nature}
Marian Bogu{\~n}{\'a}, Claudio Castellano, and Romualdo Pastor-Satorras.
\newblock Nature of the epidemic threshold for the
  susceptible-infected-susceptible dynamics in networks.
\newblock {\em Physical review letters}, 111(6):068701, 2013.

\bibitem{mata2013pair}
Ang{\'e}lica~S Mata and Silvio~C Ferreira.
\newblock Pair quenched mean-field theory for the
  susceptible-infected-susceptible model on complex networks.
\newblock {\em EPL (Europhysics Letters)}, 103(4):48003, 2013.

\bibitem{li2012susceptible}
Cong Li, Ruud van~de Bovenkamp, and Piet Van~Mieghem.
\newblock Susceptible-infected-susceptible model: A comparison of n-intertwined
  and heterogeneous mean-field approximations.
\newblock {\em Phys. Rev. E}, 86(2):026116, 2012.

\bibitem{qu2015sis}
Bo~Qu and Huijuan Wang.
\newblock {SIS} epidemic spreading with heterogeneous infection rates.
\newblock {\em arXiv preprint arXiv:1506.07293}, 2015.

\bibitem{buono2013slow}
C~Buono, F~Vazquez, PA~Macri, and LA~Braunstein.
\newblock Slow epidemic extinction in populations with heterogeneous infection
  rates.
\newblock {\em Physical Review E}, 88(2):022813, 2013.

\bibitem{fu2008epidemic}
Xinchu Fu, Michael Small, David~M Walker, and Haifeng Zhang.
\newblock Epidemic dynamics on scale-free networks with piecewise linear
  infectivity and immunization.
\newblock {\em Phys. Rev. E}, 77(3):036113, 2008.

\bibitem{yang2012epidemic}
Zimo Yang and Tao Zhou.
\newblock Epidemic spreading in weighted networks: an edge-based mean-field
  solution.
\newblock {\em Physical Review E}, 85(5):056106, 2012.

\bibitem{qu2016heterogeneous}
Bo~Qu and Huijuan Wang.
\newblock {SIS} epidemic spreading with correlated heterogeneous infection
  rates.
\newblock {\em arXiv preprint arXiv:1608.07327}, 2016.

\bibitem{barrat2004architecture}
Alain Barrat, Marc Barthelemy, Romualdo Pastor-Satorras, and Alessandro
  Vespignani.
\newblock The architecture of complex weighted networks.
\newblock {\em Proceedings of the National Academy of Sciences of the United
  States of America}, 101(11):3747--3752, 2004.

\bibitem{macdonald2005minimum}
PJ~Macdonald, E~Almaas, and A-L Barab{\'a}si.
\newblock Minimum spanning trees of weighted scale-free networks.
\newblock {\em EPL (Europhysics Letters)}, 72(2):308, 2005.

\bibitem{li2004statistical}
Wei Li and Xu~Cai.
\newblock Statistical analysis of airport network of china.
\newblock {\em Physical Review E}, 69(4):046106, 2004.

\bibitem{caldarelli2000fractal}
Guido Caldarelli, Riccardo Marchetti, and Luciano Pietronero.
\newblock The fractal properties of internet.
\newblock {\em EPL (Europhysics Letters)}, 52(4):386, 2000.

\bibitem{albert1999internet}
R{\'e}ka Albert, Hawoong Jeong, and Albert-L{\'a}szl{\'o} Barab{\'a}si.
\newblock Internet: Diameter of the world-wide web.
\newblock {\em Nature}, 401(6749):130--131, 1999.

\bibitem{barabasi1999emergence}
Albert-L{\'a}szl{\'o} Barab{\'a}si and R{\'e}ka Albert.
\newblock Emergence of scaling in random networks.
\newblock {\em Science}, 286(5439):509--512, 1999.

\bibitem{PhysRevLett.85.4626}
Reuven Cohen, Keren Erez, Daniel ben Avraham, and Shlomo Havlin.
\newblock Resilience of the internet to random breakdowns.
\newblock {\em Physical Review Letters}, 85:4626--4628, Nov 2000.

\bibitem{erdds1959random}
Paul Erd{\H{o}}s and Alfr{\'e}d R{\'e}nyi.
\newblock On random graphs i.
\newblock {\em Publ. Math. Debrecen}, 6:290--297, 1959.

\bibitem{van2006performance}
Piet Van~Mieghem.
\newblock {\em Performance analysis of communications networks and systems}.
\newblock Cambridge University Press, 2014.

\bibitem{WANGWenBin:2143}
Wenbin Wang, Ziniu Wu, Chunfeng Wang, and Ruifeng Hu.
\newblock Modelling the spreading rate of controlled communicable epidemics
  through an entropy-based thermodynamic model.
\newblock {\em Sci. Sin.-Phys. Mech. Astron.}, 56(11):2143, 2013.

\end{thebibliography}
\end{document}